\documentclass{emulateapj}
\def\cjaa{Chinese J. Astron. Astrophys.}

\received{2003 December 17}
\begin{document}

\title{X-ray triple rings around the M87 jets in the central Virgo cluster}

\author{
Hua Feng\altaffilmark{1}, Shuang-Nan Zhang\altaffilmark{2-5},
Yu-Qing Lou\altaffilmark{2,6,7} and Ti-Pei Li\altaffilmark{2,4}}

\altaffiltext{1}{Department of Engineering Physics and Center for
Astrophysics, Tsinghua University, Beijing 100084, China}

\altaffiltext{2}{Physics Department and Center for Astrophysics,
Tsinghua University, Beijing 100084, China}

\altaffiltext{3}{Physics Department, University of Alabama
in Huntsville, Huntsville, AL 35899, USA}

\altaffiltext{4}{Laboratory for Particle Astrophysics,
Institute of High Energy Physics, Chinese Academy of
Sciences, Beijing 100039, China}

\altaffiltext{5}{Space Science Laboratory, NASA Marshall
Space Flight Center, SD50, Huntsville, AL 35812, USA}

\altaffiltext{6}{Department of Astronomy and Astrophysics,
The University of Chicago, 5640 South Ellis Avenue, IL 60637, USA}

\altaffiltext{7}{National Astronomical Observatories, Chinese
Academy of Sciences, A20, Datun Road, Beijing 100012, China}

\shortauthors{Feng, Zhang, Lou, Li}

\shorttitle{X-ray Triple-Ring Structure of M87}

\begin{abstract}
The \textit{Chandra} X-ray data of the central Virgo cluster are
re-examined to reveal a triple-ring structure around the galaxy
M87, reminiscent of the spectacular triple-ring pattern of the
SN1987A in the Large Magellanic Cloud (LMC). In the sky plane, the
two apparent smaller ellipses are roughly aligned along the M87
jets; the larger ring centers at the M87 nucleus and is likely a
circle roughly perpendicular to the M87 jet. Certain similarities
of these two triple-ring structures might hint at similar
processes that operate in these two systems with entirely
different sizes and mass scales. We suspect that a major merging
event of two galaxies with nuclear supermassive black holes
(SMBHs) might create such a triple-ring structure and drove
acoustic and internal gravity waves far and near. The M87 jets are
perhaps powered by a spinning SMBH resulting from this
catastrophic merging event.
\end{abstract}


\section{Introduction}

Many similar astrophysical phenomena happen on totally different
spatial and temporal scales or energy and mass scales. Examples
include collimated jets from active galactic nuclei
(AGNs) containing SMBHs \citep{beg03}, from microquasars
containing stellar mass black holes \citep{mir99} and from young
stellar objects \citep{ray96}; dynamical roles of waves in spiral
galaxies \citep{lin64,fan96,lou98}, in planetary rings
\citep{gt78} and in galaxy clusters \citep{fab03a}; similar
high-temperature atmospheres of the Sun and of stellar mass black
hole systems \citep{zha00}; similar gamma-ray flashes and bursts
from explosions of perhaps massive stars at cosmological distances
\citep{mes01}, from solar flares \citep{hai91} and from the
Earth's atmosphere \citep{fis94,fen02}. We report here the
detection of an X-ray triple-ring structure in the core of the
Virgo galaxy cluster, most likely associated with the AGN of the
galaxy M87 (NGC 4486) and its powerful jet. This triple-ring
structure is reminiscent of the spectacular triple-ring pattern of
the supernova 1987A in the LMC \citep{bur95}.

M87 is an active galaxy with relativistic jets in the central
Virgo cluster. The X-ray morphology and spectroscopy of the Virgo
cluster have been studied previously using data from
\textit{Chandra} \citep{you02} and \textit{XMM-Newton}
\citep{boh01,bel01}. Combining \textit{Chandra}, \textit{XMM} and
\textit{ROSAT} data, \citet[hereafter F04]{for04} studied the
Virgo cluster on various scales to identify cooling flow quenching
by an AGN energy input. They noticed two prolate features around
the M87 jet and counter jet, and referred to them as ``cavities''
caused by plasma expansions.

In this Letter, we re-analyze the \textit{Chandra} data and focus
on the core region around M87. Two smaller ring structures are
identified to encircle the ``cavities'' noticed by F04; a third
larger ring centers at the M87 nucleus. We adopt the distance to
M87 as $\sim 16$ Mpc with 1\arcsec\ for $\sim 78$ pc in the image
\citep{whi95}.

\section{Data acquisition and analysis }

We process the data of two observations (Obs-ID 2707 and 352)
pointed at M87 in the central Virgo cluster by the ACIS-S
(Advanced CCD Imaging Spectrometer Spectroscopic array) instrument
aboard NASA's \textit{Chandra X-ray Observatory} for a total
effective exposure of 119.2 ks. The data are screened for flares
where count rates are at least 3-$\sigma$ away from the mean rate
in the S1 chip for 2.5--6.0 keV. The exposures before/after
screening are 98.7/89.5 ks for Obs-ID 2707 and 37.7/29.7 ks for
Obs-ID 352. These two screened data sets are merged using the
\textit{merge\_all} script in \textit{CIAO} 3.0.1, and further
corrected with an exposure map. Another two long observations of
M87 (Obs-ID 3717 and 1808) are not included for analysis because
of serious flare contamination.

We avoid the over-brightness of M87 and the jet by extracting the
normalized brightness in the range [0, 8\%] and renormalizing
linearly to [0, 1], i.e., let $B_1$ and $B_2$ be the brightness
maps before and after the adjustment, we take $0\leq B_1\leq0.08$
from a normalized brightness map and renormalize it with
$B_2=B_1/0.08$. The 0.5--2.5 keV image of the central Virgo
cluster is displayed in Figure \ref{virgo}a by a Gaussian
smoothing of 1.5\arcsec\ FWHM. To reveal structures of smaller
scales in the image, we apply an unsharp mask. A smoothed image
$G$ is obtained by convoluting the raw image $I$ with a Gaussian
function. The sharpened image $U$ is a weighted subtraction of $I$
and $G$, viz. $U=I+a(I-G)$ where $a$ is a constant. By trials, we
choose a Gaussian function of 15\arcsec\ FWHM with $a=500\%$. The
resulting image of Figure \ref{virgo}b is obtained by smoothing
the sharpened image with a Gaussian function of 1.5\arcsec\ FWHM
and by adjusting the contrast to enhance visibility of the
triple-ring feature; the contrast enhancement is done by
displaying only the image within the brightness range of
15.5--24\%, i.e., 1.24--1.92\% in the original image.

For spectra along the rings, a hardness ratio map is shown in
Figure \ref{hr} for the central Virgo cluster where the hardness
ratio is defined as $H=(c_2-c_1)/(c_2+c_1)$ with $c_1$ and $c_2$
for photon counts in the lower and higher energy bands,
respectively \citep{li01}. We select 0.5--1.0 keV and 1.0--2.5 keV
as the lower and higher bands. The images are smoothed with a
2\arcsec\ FWHM Gaussian function before creating a hardness ratio
map in Figure \ref{hr}.

\section{The triple-ring structure}

\begin{figure}[t]
  \centering
  \plotone{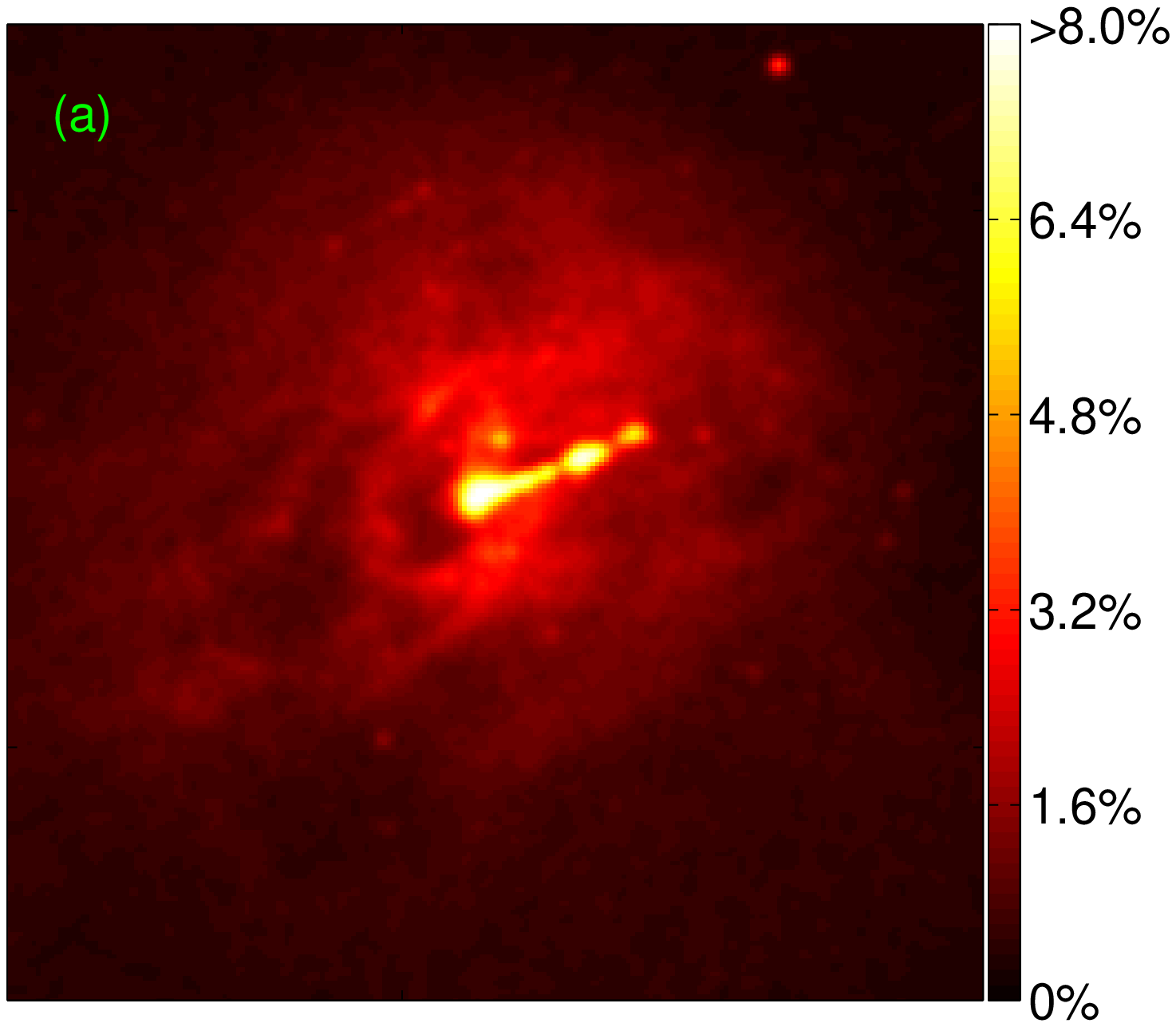}\\\plotone{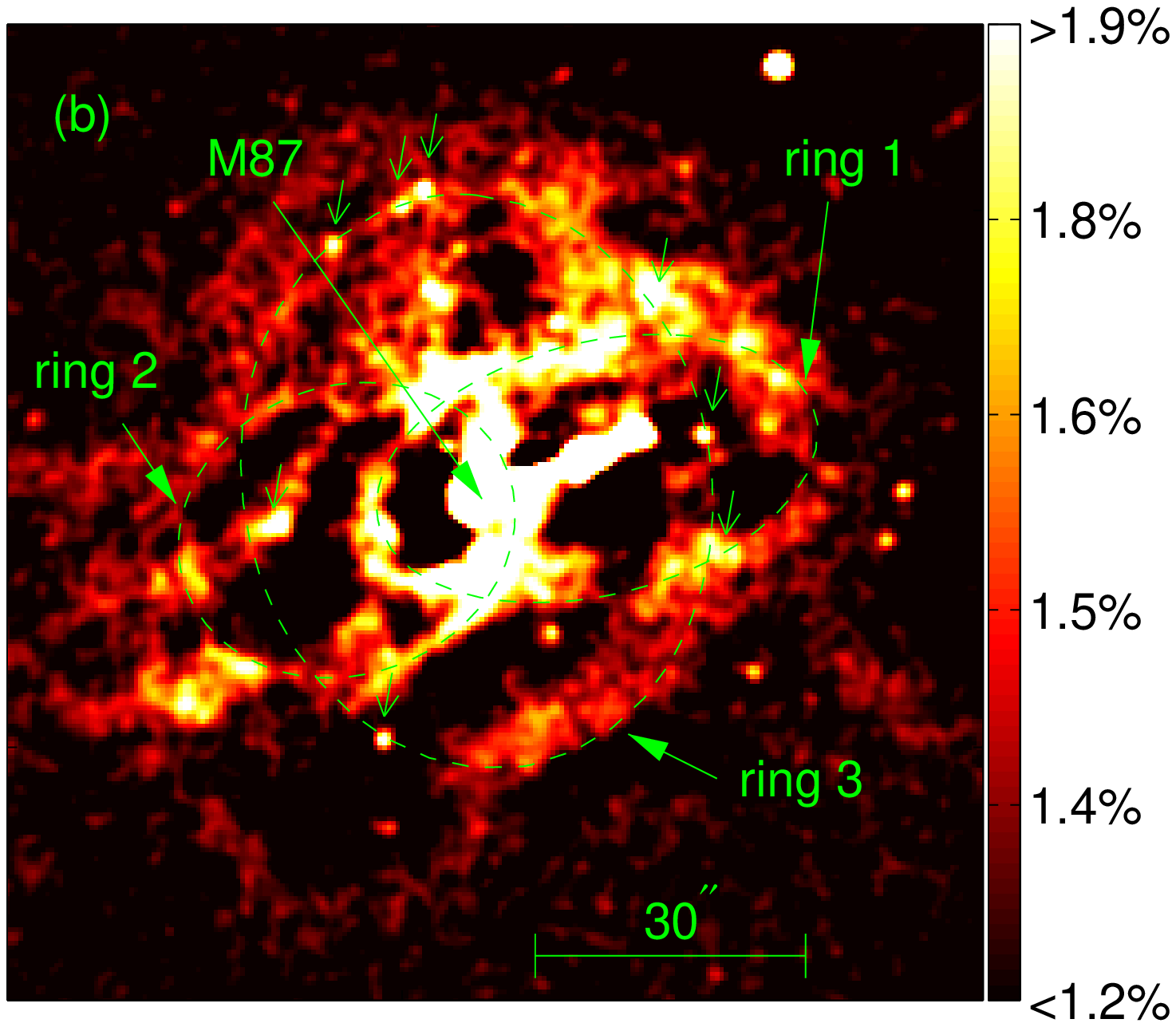}
  \caption{Two \textit{Chandra} 0.5--2.5 keV X-ray images of the
central Virgo cluster with color bars indicating the fractional
flux. (a) --- the core image plotted in linear intensity scale,
smoothed by a Gaussian function of 1.5\arcsec\ FWHM and with a
contrast adjustment to avoid the over-brightness of M87 nucleus
and jets; (b) --- the triple-ring feature (dashes) enhanced image
with an unsharp mask. In the sharpened image, two rings (northwest
ring 1 and southeast ring 2) are roughly along the M87 jet. The
third ring 3 can be seen in broken segments, viz. the northern and
southern portions fit in an ellipse with the M87 nucleus as its
center. Ring 3 passes across eight bright X-ray knots identified
by small open arrows.}
  \label{virgo}
\end{figure}

Two nested X-ray ellipses (dashed lines marked as ring 1 and 2 in
Figure \ref{virgo}b) are revealed around the M87 jets. A third
larger ellipse can be partially seen (dashed line marked as ring 3
in Figure \ref{virgo}b). The reality of ring 1 and 2 is evident by
Figure \ref{virgo}b. Besides the northern and southern diffuse
arcs for the initial identification of ring 3, the ring also
passes through eight bright X-ray knots (marked by small open
arrows) that lend further support for its reality. The X-ray
morphology of the Virgo cluster is fairly complex with structures
on scales from $\sim 1-50$ kpc (F04). We mainly focus on the
triple-ring structure.

Each presumed circular ring fits a projected ellipse using five
parameters (see Table~\ref{rings_tab}), viz., the right ascension
(RA) and declination (Dec), semi major and minor axes, and the
inclination angle of the semi major axis to the local hour circle
line. Because the three X-ray rings can not be readily recognized
by our software, errors in these fitting parameters are estimated
empirically. We note that position uncertainties are $\lesssim
1-2\arcsec$ ($\sim 0.1$ kpc) and angular uncertainties are
$\lesssim 5\arcdeg$.

\begin{deluxetable}{lccccccc}[t]
\tablecaption{Fitting parameters and orientations of the triple
rings\tablenotemark{1} \label{rings_tab}}

\tablehead{
 \colhead{} & \multicolumn{4}{c}{projected ellipses} & \multicolumn{3}{c}{reconstructed rings}\\
 \cline{2-4}  \cline{6-8}\\
 \colhead{object} & \colhead{center (RA, Dec)} & \colhead{a} & \colhead{b} & \colhead{$\theta$} & \colhead{$\alpha$} & \colhead{$\Phi$} & \colhead{$\Phi'$}\\
 \colhead{} & \colhead{(\arcdeg)} & \colhead{(kpc)} & \colhead{(kpc)} & \colhead{(\arcdeg)} & \colhead{(\arcdeg)} & \colhead{(\arcdeg)} & \colhead{(\arcdeg)}
}
 \startdata
ring 1 & 187.7021, 12.3920 & 1.9 & 1.1 & 76 & 55 & 61 & 55\\
ring 2 & 187.7100, 12.3901 & 1.5 & 1.3 & 70 & 30 & 51 & 33\\
ring 3 & 187.7059, 12.3916 & 2.5 & 2.0 &-10 & 37 & 9  & 22\\
nucleus & 187.7059, 12.3911 & \nodata & \nodata & \nodata & \nodata & \nodata & \nodata \\
jet & \nodata & \nodata & \nodata & 70 & 43 & 0 & 0\\
 \enddata
 \tablenotetext{1}{a: semi major axis; b: semi minor axis; $\theta$:
Angle from the major axis (or M87 jet direction) to the local hour
circle line; $\alpha$: angle from the ring normal (first 3 rows)
or M87 jet direction (the last row) to the LOS; $\Phi$: angle
between the ring normal and the M87 jet (43\arcdeg\ to LOS);
$\Phi'$: same with $\Phi$ but taken the jet orientation as
15\arcdeg\ to LOS.}
\end{deluxetable}

It is possible that the observed ellipses are projections of
circular rings in the sky plane. We reconstruct 3-D orientations
of the three circular rings. The jet orientation may be taken as
either 43\arcdeg\ \citep{bir95} relative to the line of sight
(LOS) away from us or more likely 15\arcdeg\ \citep[within
19\arcdeg\ as inferred by][]{bir99}. The angle of the projected
jet to the local hour circle line is $\sim$ 70\arcdeg\ (Figure
\ref{virgo}). Thus, the angle $\Phi$ (see Table~\ref{rings_tab})
of the reconstructed circular ring normal to the jet direction (if
43\arcdeg\ to LOS) is 61\arcdeg, 51\arcdeg\ and 9\arcdeg\ for ring
1, 2 and 3, respectively. This means that ring 3 is likely a
circular ring nearly perpendicular to the M87 jet. For a jet
orientation of 15\arcdeg\ to the LOS, the corresponding angles
$\Phi'$ of ring normal to the jet (see Table~\ref{rings_tab}) are
55\arcdeg, 33\arcdeg\ and 22\arcdeg, respectively. The derived
values of $\Phi$ and $\Phi'$ for presumed `circular' ring 1 and 2
are fairly large; these two rings might be elliptical in the 3-D
space or their orientations are not perpendicular to the jet. Our
estimate excludes the possibility that the elliptical shape
of rings 1 and 2 is caused by relativistic projection effects if
the two circular rings expand radially away from the jet axis
while moving apart from each other relativistically along the line
connecting the two ring centers, because by special relativity
the ring would be elongated transverse to the jet
direction. We note that ring 3 passes through eight X-ray bright
knots with its center located exactly at the M87 nucleus. The
chance of such a coincidence is small.

As projected ring 1 and 2 roughly align along the direction of the
M87 jet and counter jet, their intrinsic elliptical (instead of
circular) shapes in 3-D space are plausible. Given relatively
small angle of the ring 3 normal to the jet (9\arcdeg\ or
22\arcdeg), one may qualitatively regard ring 3 as circular and
perpendicular to the jet.

\section{Spectroscopy of the three rings}
\label{sec:spec}

\begin{figure}[t]
  \plotone{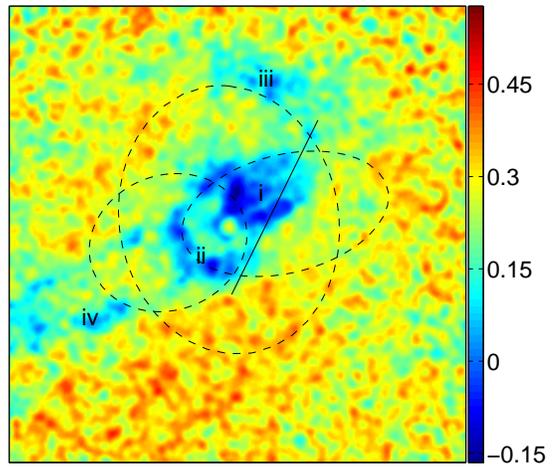}
  \caption{Hardness ratio map in the central
Virgo cluster. The two energy bands are 0.5--1.0 keV and 1.0--2.5
keV. Each pixel stands for $H=(c_2-c_1) /(c_2+c_1)$ where $c_1$
and $c_2$ are photon counts at this pixel in the lower and higher
energy bands respectively. Four softest regions are labeled. The
solid line indicates a separation of the two parts along ring 1
for spectral comparisons (see \S\ \ref{sec:spec}).}
  \label{hr}
\end{figure}

From Figure \ref{hr}, the softest regions locate at (i) around the
jet within a $\sim$20\arcsec\ radius sector from west to north of
the nucleus, (ii) an extended arc coincident with the east portion
of ring 1 from north to south of the nucleus, (iii) a short arc
coincident with ring 3 about $\sim$30\arcsec\ away from the
nucleus in the northwest and (iv) some extended filaments
southeast of the nucleus outside the three rings. By comparing
Figure \ref{hr} with the H$\alpha$+[N {\footnotesize II}]
emissions of the central Virgo cluster \citep{spa93}, we find that
the soft region (i) and (iv) are correlated with H$\alpha$+[N
{\footnotesize II}] filaments. Similar correlations were also
found in the Perseus cluster \citep{fab03b}, indicating that
intracluster gases become cooler when adjacent to the filaments
due to thermal conduction.

The association of regions (ii) and (iii) with ring 1 and 3
respectively is interesting with region (ii) being more striking.
A harder spectrum may correspond to a higher temperature.
Different temperatures along ring 1 indicate that X-ray emissions
may not be simply caused by a high density of electrons alone, but
may involve supersonic shock flows from the AGN. Other parts of
rings without temperature difference relative to surroundings
might indicate subsonic flows.

Energy spectra are derived from two segments of ring 1, viz., the
east part identified with region (ii) and the west part. To
compare the spectra in these two parts with their environments,
the source region is partitioned as an elliptical annulus along
ring 1 with a width of $\sim$4\arcsec\ and a background region of
a bigger elliptical annulus outside 5\arcsec\ of ring 1 with a
width of $\sim$2\arcsec. The source and background regions are
separated into two parts by a straight solid line in Figure
\ref{hr}.

In Figure \ref{hr}, the east part carries features, while the west
part blends with the environment. We adopt a hot diffuse gas
emission model (VMEKAL) to fit the spectra of both parts, with
free parameters of equivalent hydrogen column, temperature and
elemental abundances of O, Ne, Mg, Si, S and Fe. The fitting is
carried out using \textit{XSPEC} 11.3. The best-fit temperatures
are $0.74_{-0.05}^{+0.04}$ keV ($\chi^2/\mathrm{d.o.f.}=71/54$)
for the east part and $1.32_{-0.05}^{+0.04}$ keV
($\chi^2/\mathrm{d.o.f.}=53/55$) for the west part. A remarkable
difference in plasma temperature can be seen from energy spectra
using the hardness ratio map in Figure \ref{hr}. In contrast, only
the central ring of SN1987A can be resolved by \textit{Chandra}
using a sub-pixel technique \citep{bur00,par02,mic02}. As its
X-ray spectrum may be fit by a plane-parallel shock model, the
central ring of SN1987A is thought to involve an expanding blast
wave. By this clue, we attempt to fit the spectra of both east and
west parts of ring 1 (Figure \ref{virgo}b) using the
plane-parallel shock model (VPSHOCK) with the same free parameters
of elemental abundance indicated above. The west part spectrum
does not fit the plane-parallel model well with a large
$\chi^2/\mathrm{d.o.f.}\sim 5$, while the east part spectrum fits
well with the shock model. The best-fit temperature is
$0.71_{-0.04}^{+0.04}$ keV ($\chi^2/\mathrm{d.o.f.}=77/53$),
consistent with the hot plasma model. This suggests that ring 1
may involve a blast wave where the west part might have been
thermalized with the surroundings and the east part might be in
the process of being thermalized. For the east part, the best-fit
abundances of elements O, Si, S, and Fe in the thermal model are
systematically lower than that in the west part. The enhancement
of abundances in the warmer region may indicate more active
nuclear processes in the past, perhaps caused by the
thermalization in the west part. The similar temperature-abundance
correlation is also found with \textit{XMM} in the inner region
for radius $\lesssim$1\arcmin\ \citep{boh01} and for the two
large-scale arms \citep{bel01}. Because of uncertainties in
estimating abundances with \textit{Chandra} data for the fine
structures, we refrain from more detailed comparisons with
previous results of \textit{XMM} for the larger structures.

\section{Discussions}

The triple-ring pattern revealed in the central Virgo cluster is
the only one observed since the discovery of triple rings in
SN1987A\footnote{http://oposite.stsci.edu/pubinfo/jpeg/SN1987A\_Rings.jpg}
\citep{bur95}. The ring sizes in the Virgo cluster are several
thousand times larger than those of SN1987A. The triple-ring
structure of SN1987A has been simulated in terms of interacting
winds near the end of stellar evolution. According to
\citet{Tan02}, during the star's red supergiant phase, slower
winds persist. Prior to the supernova explosion, the red
supergiant somehow evolved into a blue supergiant that drove
faster winds. Magnetohydrodynamic (MHD) interactions of fast and
slow winds from a rotating star produce the triple-ring structure
in SN1987A. We suspect that a triple-ring structure might be
generic in a rotating system involving magnetized faster winds
catching up slower winds \citep{lou94,lou96}. Meanwhile a central
source of radiation is required for the rings to shine in the
X-ray band. For SN1987A, the strong UV radiation from the
supernova provides the necessary illumination.

We suspect that a catastrophic merging event around the M87
nucleus might be responsible for the triple-ring structure
revealed here. Merging of two SMBH systems has been proposed to
drive the observed X-shaped radio lobes \citep{mer02}. In this
scenario, a `slower wind' was present before the merging begins,
e.g., a mixture of galactic winds from two merging galaxies. A
`faster wind' was then driven during the merging process. It has
been suggested that an active AGN phase occurred in the core
region of M87 at about $10^8$ years ago \citep{kai03}; perhaps,
this AGN was triggered by the merging event discussed above,
when significantly higher accretion rate was provided by the
merger. The merger may fuel both the radiation energy output of
the AGN and the growth of the supermassive black hole in the
center; huge amount of high energy radiation shines the triple
rings in X-ray bands. It is also plausible that the resulting
SMBH spins rapidly to power the highly collimated M87 jets
\citep{bla77}.

F04 revealed two expanding cavities around the M87 jet and counter
jet, enclosed within our ring 1 and 2 respectively. By different
external pressures, they inferred the far front of the cavities is
mildly supersonic while the inner part, close to the nucleus, is
collapsing. We suggest that these two rings are actual rings
instead of projections of prolate spheres, because when two
cavities collide into each other it is hard to maintain their
spherical shapes in the spatially overlapping portions of the two
spheroids, and consequently their projections cannot be viewed as
two complete and nested rings. We further suggest that when the
jets form, they push materials around and drive powerful blast
waves of ring shapes.

Qualitatively similar central structures are present in the
clusters Perseus \citep{fab03a}, Hydra A \citep{mcn01}, Centaurus
A \citep{kra02,kar02} and Virgo (F04). Physically, core activities
can excite large-scale acoustic waves ($p-$modes) and internal
gravity waves ($g-$modes) in a deep gravitational potential well
of central galaxy cluster \citep{lou95}. Cluster core activities
can be either violent and sporadic or sustained at a level or a
combination of both. Depending on the phase of acoustic wave
propagation and on physical properties of core environment, we may
observe various X-ray morphologies superposed onto a smooth
luminous core. Beautiful acoustic rings or arcs have been revealed
in Perseus \citep{fab03a} and in Virgo on larger scales (F04).
Intense core X-ray emissions tend to induce inflows of gas that
perturb the dark matter via gravity. The violent relaxation
proceeds in a sound-crossing timescale \citep{lb67}. This could be
a persistent source of $p-$modes and $g-$modes in a cluster core.
By such wave and Laudau-damping processes, slowly inward drifting
gas and dark matter are virialized to sustain X-ray losses from
hot electron gas. As a cluster core may involve magnetic field of
strengths up to $\sim 30-40\mu$G, MHD waves should also play
significant roles. Based on the hardness ratio map and the energy
spectra fitting, we infer only parts of ring 1 and 3 involve
supersonic shock flows and are in processes of being thermalized
with the surroundings. By this reasoning, other parts without
apparent difference with surroundings might have been thermalized
already. In summary, we hope that the triple-ring structure of M87
in the central Virgo cluster would stimulate more theoretical,
simulation and observational studies of MHD flow, wave and shock
interactions.

\acknowledgments We thank the anonymous referee whose insightful
and instructive comments have improved our work significantly.
This work was supported in part by the Special Funds for Major
State Basic Science Research Projects, by the Directional Research
Project on High Energy Astrophysics of the Chinese Academy of
Sciences, and by the National Natural Science Foundation of China
(NSFC) grants No. 10233030, 10028306 and 10373009. SNZ also
acknowledges supports by NASA's Marshall Space Flight Center and
by NASA's Long Term Space Astrophysics Program. YQL also
acknowledges partial support by the ASCI Center for Astrophysical
Thermonuclear Flashes at the U. of Chicago under DOE contract
B341495, and by the Yangtze Endowment from the MOE through the
Tsinghua Univ. The data were taken from the HEASARC online service
of the NASA/GSFC.

\clearpage

\clearpage

\end{document}